\documentstyle[12pt,fleqn]{article}
\def\x0{\xi_0}
\def\cm{{\rm cm^{-3}}}
\def\hcm{{\rm cm^{-3}}}
\def\dd{D_{10}}
\def\vv{v_{220}}
\def\d101{D_{10}^{-1}}
\def\v2202{v_{220}^2}
\def\be{\begin{equation}}
\def\ee{\end{equation}}
\def\la{\mathrel{\mathpalette\fun <}}
\def\ga{\mathrel{\mathpalette\fun >}}
\def\fun#1#2{\lower3.6pt\vbox{\baselineskip0pt\lineskip.9pt
        \ialign{$\mathsurround=0pt#1\hfill##\hfil$\crcr#2\crcr\sim\crcr}}}

\begin{document}
\begin{titlepage}
\null\vspace{-62pt}
\vspace{1.5in}
\baselineskip 12pt
\centerline{{\large \bf ~~Hydrogen Clouds and the MACHO/EROS Events}}
\vspace{0.2in}
\centerline{~~RICHARD N. HENRIKSEN\footnote{E-mail address:
henriksen@bill.Phy.Queensu.CA} and
LAWRENCE M. WIDROW\footnote{E-mail address:
widrow@lola.Phy.Queensu.CA}}
\vspace{0.2in}
\centerline{{\it ~~Department of Physics}}
\centerline{{\it ~~Queen's University, Kingston, Ontario, K7L 3N6, CANADA}}
\vspace{.5in}
\baselineskip=12pt
\centerline{ABSTRACT}
\baselineskip=12pt
\vspace{.5in}
We propose that the recently reported MACHO/EROS events correspond to
gravitational amplification by dark clouds rather than compact objects.  These
clouds must be very dense with $M\sim 0.1 ~M_\odot$ and $R\la
10^{14}~{\rm cm}$.  In all likelihood,
the clouds will be members of a family of objects with different sizes and
masses.  We therefore
expect events of longer duration than the ones reported by the MACHO and EROS
groups but with light curves very different from the ones derived assuming
point mass lenses.  We suggest that one such event has already been observed in
radio measurements of the quasar 1502+106. The abundances of free electrons,
metals, complex molecules, and dust grains are constrained to be very small
suggesting that the clouds are formed from a primordial mixture of hydrogen and
helium.
Cosmic rays and background
UV radiation ionize a halo around the cloud.
Radio waves from distant sources will be scattered by the electrons
in this halo, an effect which may have already been observed
in quasars such as 1502+106.  We argue
that dark clouds are a viable alternative to compact objects for baryonic
dark matter in the halo.
\end{titlepage}
\newpage

\baselineskip=18pt

\vspace{24pt}

\centerline{\bf I.~Introduction}
\bigskip
Recently, the MACHO (Alcock 1993)
and EROS (Aubourg 1993) collaborations announced that
gravitational amplification or
microlensing of light from stars in the Large Magellanic Cloud
may have been observed.
Three `events' were reported where an
event corresponds to a transient, time-symmetric,
achromatic brightening of an LMC star (Paczynski 1986).
The interpretation of these events is that a compact object has passed very
close to the line of sight to the star.  The masses inferred from the three
events are all in the range $0.03-0.3~M_\odot$.
This is consistent with the lenses being
brown dwarfs or low-mass stellar remnants though more exotic objects such as
black holes can also explain the observations.
Microlensing events are extremely rare and some five million stars
were observed in order to find
just these three events.  Still, the observations
suggest that some, and perhaps a significant fraction of dark matter
in our galaxy is in
the form of nonluminous compact objects and therefore, in all likelihood,
dark baryons.
This conclusion is consistent with results from primordial nucleosynthesis,
which indicate that the density in baryons is greater than the
density in luminous matter (Walker et al.~1991).  In any case,
the MACHO/EROS discoveries, if confirmed, will have
important implications for star formation, galaxy evolution, and cosmology.

In this work, we suggest that the lenses are dense clouds
rather than compact objects.
While brown dwarfs or stellar remnants are appealing
dark matter candidates (they are after all, the only dark matter candidates
known to exist) the discovery that stellar-type compact objects
make up a {\it significant} fraction of the
halo dark matter would be surprising from the standpoint of both
theory and observation (see, for example, Liebert and Probst 1987).
Dark baryon clouds represent an alternative way
of hiding large amounts of baryons (Pfenniger, Combes, and
Martinet 1993 (PCM), Pfenniger and Combes 1993 (PC)).

In Section II we derive light curves for a number of extended lens models
illustrating that a variety of clouds can explain the MACHO and EROS events.
In Section III we discuss observational and theoretical evidence for
baryonic clouds.  Our focus from the observational side is to
connect the lensing events with `Extreme Scattering Events'
observed in radio measurements of quasars (Fiedler et al.~1987 (F1),
Fiedler et al.~1993 (F2)).  Our theoretical discussion attempts to
make plausible the idea that dense baryon clouds
are a long-lived, if not final phase in the evolution of
primordial baryonic matter.  It also indicates that the clouds will be
distributed in mass and size suggesting observations which can test the
theory.

\bigskip
\centerline{\bf II.  Lensing by Extended Objects}
\bigskip

Consider a compact gravitational lens of mass $M$ at distance $l$
from the observer and a
source at distance $L$.  Significant
amplification occurs when the lens passes within
a distance $\xi_0\equiv\sqrt{4GMD/c^2}$ of the line of sight to the
source where $D\equiv l\left (L-l\right )/L$.
$\xi_0=4\times 10^{13} \,{\rm cm}\left (M/M_\odot\right )^{1/2}
\left (D/ {\rm kpc}\right )^{1/2}$ is often referred to as the Einstein radius.
For a source in the LMC and a $0.1~M_\odot$ lens in the halo of our galaxy,
$D\la 13~{\rm kpc}$ and $\xi_0$ corresponds to roughly
$10^{-4}~{\rm arcsec}$ on the sky.  An extended object
with a characteristic size $R\ll\xi_0$ will behave much like
a point mass lens while an object with $R\gg\xi_0$ will not cause
significant amplification.  It is the intermediate regime that interests us
here.

Now consider a few simple examples of extended lenses (see, for example,
Schneider et al., 1992, and references therein).  The lenses are assumed to
be transparent at the frequencies where lensing is observed.  Furthermore, for
simplicity, only lenses that are axially
symmetric about the line connecting the center of the lens to the observer
are considered.
Let $\Sigma(\xi)=\Sigma_{\rm cr}\kappa(x)$ be the surface density where
$\xi=\xi_0x$ is the distance from the center of the lens
in the lens plane and $\Sigma_{\rm cr}\equiv
c^2/4\pi GD$ defines the critical surface density.
The lens equation,
\be
\label{eq:LENS}
y~=~x~-~\frac{m(x)}{x}~,
\ee
determines where the light ray from the source crosses the lens plane.
Here $m(x)=2\int_0^x x' dx' \kappa(x')$ is the (dimensionless) mass within a
circle of radius $x$ and $y=\eta/\eta_0$ is the (dimensionless) position of
the source in the source plane where $\eta_0=\x0 L/l$.
In general, there can be more than one solution to
Eq.~(\ref{eq:LENS}) corresponding to multiple images of the source.  For each
image $i$, the amplification is given by
\be
A_i=\left (\frac{y}{x}\frac{dy}{dx}\right )^{-1}_i~.
\ee
Microlensing refers to the case where the images cannot be resolved.  What is
measured is then the total amplification, $A \equiv \Sigma_i A_i$.

For the point mass lens, $m(x)=1$, there are two images,
and
\be
A = { y^2 + 2 \over y \sqrt{y^2+4} }~.
\ee
Light curves are constructed by setting $y=\left (y^2_0+\tau^2\right )^{1/2}$
where $\tau=vt/\xi_0$, $t$ is time, and $v$ is the transverse velocity
of the lens assuming source and observer are fixed (Paczynski 1986).
(See also Griest (1991) for
a more general discussion where the velocities of observer, lens, and source
are taken into account.)

A particularly simple model for an extended lens is the
singular isothermal sphere (SIS).  Here,
$\rho(r)=\sigma^2/2\pi Gr^2$, $\Sigma(\xi)
=\sigma^2/2G\xi$ and $\xi_0=4\pi\left (\sigma/c\right )^2 D$
where $\sigma$ is the line-of-sight velocity dispersion.
Of course real clouds have a large radius cut-off
so that the total mass is finite, and an inner core radius which removes the
singularity at $r=0$.  Still, the SIS works fine as a model
for realistic isothermal spheres so long as the
light ray from the source
crosses the lens plane outside the core radius and inside the cut-off
radius.  There are either one or two images depending on the
value of $y$ and the total amplification is
\be
A=\cases{2/y&0$\le y\le 1$\cr
	\left (1+y\right )/y&$1\le y$\cr}~~.
\ee
Again, light curves are found by setting $y=\left (y_0^2+\tau^2\right )^{1/2}$.
In Fig.\,1a we compare light curves for
the point mass and SIS lenses.  The curves are chosen to match the
MACHO group event (Alcock 1993).  In particular, $y_0$ is chosen for each model
so that the peak amplitude is $A_{\rm max}=6.86$.  The horizontal scale is
set so that the peak occurs at $\tau=0$ and $A=2.0$ at $\tau=\pm 1$.
For the MACHO event, $A=2$ occurs at $t\simeq\pm 9~{\rm days}$ from the peak.
Using this we find that $M=0.06\, M_\odot \d101\v2202$
for the point mass lens and
$\sigma=2~{\rm km/sec}\,\d101\v2202$
for the SIS lens where $D=\dd 10~{\rm kpc}$ and $v=\vv 220 \,{\rm km/sec}$.

A second model is a spheroid of constant density $\rho_0$, and radius $R=\left
(3M/4\pi\rho_0\right )^{1/3}\equiv\xi_0 x_0$.  In this case, there are one,
two, or
three images depending on the values of $y$ and $x_0$.  The lens equation is
solved numerically and one can once again generate light curves
to match the observations.  Fig.\,1b for example, compares
theoretical light curves for a spheroid lens with $x_0=0.6$ and
$M=0.04\,M_\odot\,\d101\v2202$
and the point mass lens of Fig.~1a.  The cusps occur where the
number of images changes from $1$ to $3$ and the amplification formally becomes
infinite though not if we take the finite extent of the source into account.

Our finally example is a disc of radius $R=\xi_0x_0$ with constant
surface density
$\Sigma(x)=M/\pi R^2=\Sigma_{\rm cr}/x_0^2$.  The model is meant to correspond
to an object flattened due to rotation and viewed face on.  Again, there are
one, two, or three images
depending on the values of $y$ and $x_0$.  In order to fit the MACHO event
we require that $x_0<1$.  For this case
\be
A=\cases{
\frac{y^2+2}{y\sqrt{y^2+4}}+
\left (\frac{x_0^2}{1-x_0^2}\right )^2
&~~for~~ $0\le y\le \frac{1-x_0^2}{x_0}$\cr
\frac{y^2+2+y\sqrt{y^2+4}}{2y\sqrt{y^2+4}}&~~for~~ $\frac{1-x_0^2}{x_0}\le
y$\cr}
\ee
Fig.\,1c compares the disc model ($x_0=0.6,~M=0.05\,M_\odot\d101\v2202$)
and the point mass lens discussed above.

One should keep in mind that the data for the MACHO and EROS events consists of
10-20 measurements for each light curve with error bars varying from
point to point and typically in the range $|\Delta A|\sim 0.1-1.0$.  Clearly,
these measurements are not detailed enough to distinguish among the models;
the
error bars are too large to rule out, say the SIS and the sampling rate is not
high
enough to pick up the cusps which occur, for example,
in the spherical model.  However, {\it a single
event with better sampling and smaller error bars would be able to confirm or
severely constrain the extended lens hypothesis}.

\bigskip

\centerline{\bf III. Baryonic Clouds : Observation and Theory}

\bigskip

Clearly extended objects can cause gravitational amplification
consistent with the MACHO and EROS events.  But what are these objects?  One
possibility is that they are virialized clouds of collisionless, exotic
particles such as axions or supersymmetric particles.
While some small-scale clumpiness in collisionless particles
is likely to occur, indications are that the objects which form are either
too low in density (Hogan and Rees 1987, Silk and Stebbins 1993)
or too small (Kolb and Tkachev 1993).

Less speculative but more interesting is the possibility that the clouds are
composed of baryonic matter.  An object of mass $0.05 ~M_\odot$ and radius
$2\times 10^{13}\, {\rm cm}$ has a density $\rho\simeq 3\times 10^{-9}\,{\rm
g/cm^3}$,
and velocities $\simeq 2{\,\rm km/sec}$.
For atomic hydrogen, this would correspond to a number density
$n_H\simeq 2\times 10^{15}\,\hcm$ and a column density
$N_H\simeq 10^{29}\,{\rm cm^{-2}}$.  Charged particles, complex molecules, and
dust grains
can all absorb light in the visible, obscuring the background source,
and so we have strict limits on the ionization fraction and molecular and dust
densities in the clouds.  For example, the ionization fraction must be less
than about $10^{-6}$.  The density of $5000\AA$ dust grains must be $\la
10^{-6}
{\rm cm^{-3}}$.

The above arguments suggest clouds formed from a primordial mixture of
hydrogen and helium.  We now discuss both the observational evidence for their
existence and theoretical evidence that they represent a long-lived phase in
the
history of primordial matter.  Direct observation of the baryonic clouds
will be difficult because they are fairly compact.  For example, a cloud
$1\,{\rm AU}\simeq 1.5\times 10^{13}\,{\rm cm}$ in size and $1\,{\rm kpc}$
away subtends $10^{-3}\,{\rm arcsec}$ on the sky.  Take, for example, the
$21\,{\rm cm}$ line.  An optically thick cloud at $500 ^\circ K$
(corresponding to a characteristic velocity of $2\,{\rm km/s}$)
produces a flux of only $\simeq 2\times 10^{-9}\,{\rm Jy}$.
Even a VLA beam of $1\,{\rm arcsec}$, which would have something
like $100-1000$ objects in the beam (depending on the distribution of
objects with size), would receive a background signal at the level
of $\mu{\rm Jy}date$, well below current detection levels.
However, the clouds may produce other effects.  Indeed,
related observations may well be the
``Extreme Scattering Events" (ESEs) seen in radio measurements
of quasars (F1,\,F2).
At $3~{\rm GHz}$, ESEs are characterized by a flat-bottomed
flux minimum bracketted by flux maxima.  In most, but not all
cases, the $8~{\rm GHz}$ flux does not show any unusual behaviour.
So far, ten ESEs have been identified with timescales ranging from
$0.2-1~{\rm yr}$.  An ESE may have also been observed in
radio observations of the millisecond pulsar
PSR 1937+21 (Cognard et al. 1993).  Here, an unusual
event is seen in both the flux density and timing measurements of the pulsar's
signal.
F1 argue that ESEs are due
to occultations by localized regions where the electron density is
$4\times 10^3\,\cm$.  These regions are typically  $ 10 ~{\rm AU}$ in size
and $1~{\rm kpc}$ away.
Scattering of a given ray of light can be by many electron clumps,
as in the statistical approach of F2, or by
a single localized region with a smoothly varying electron density,
as in Romani et al.~(1987).  Similarly, Cognard et al.~(1993) argue that
their event is due to the passage of a fairly small ($R\simeq 0.05\,{\rm AU}$)
and slow moving ($v\simeq 15\,{\rm km/sec}$) cloud with an electron density
of $n_e\simeq 250\,\cm$.

ESEs tell us something about the electron densities in the intervening
cloud.  To determine the mass of the cloud, we need to know the
ionization fraction.  Using an ionization
fraction for dense clouds of $10^{-7}$ as estimated,
for example by Genzel (1992, p.\,342), PCM and PC
find a hydrogen density of $10^{10}~\hcm$, consistent with their picture
of hydrogen clouds as galactic dark matter (see below).
These densities are too small for gravitational lensing (at least for
a $10~{\rm AU}$ object).  Here, we
imagine a much denser cloud ($n_H=10^{15}\,\hcm$) with a low ionization
fraction
in the core.  Electron densities of $4\times 10^3\,\cm$ might be indicative of
a more highly ionized halo.
Simple calculations show that the UV backgroud, for
example, will only ionize the halo of an otherwise pure atomic hydrogen cloud
with electron densities in the halo consistent with what is required for an
ESE.

Clearly, the simultaneous observation of a gravitational lensing event and
an ESE would be powerful evidence in favour of our picture.
Remarkably, inspection of the data for 1502+106
(F2 and Figs. 2a, 2b)
shows a candidate event.  F2 interpret the variability
of this source between $\sim 1985.0$ and $\sim 1987.0$ as due to
an ESE superimposed on a
flare intrinsic to the source.  The necessary coincidence is explained
in terms of the abrupt appearance of a point-like component to the source
associated with the flare which
enhances the likelihood of an ESE.  Our interpretation is that
a super-dense cloud having a core density of $n_H\simeq 10^{15}\,\cm$
and mean ionization fraction of
$\sim 10^{-11}-10^{-12}$ has passed close to the line-of-sight to the
source.  The light curve will be affected by gravitational amplification,
electron scattering, and the finite extent of the source.
In general, the emitting region of the source is larger at longer
wavelengths.  This, together with the fact that electron scattering
is also greater at longer wavelengths implies that the $8\, {\rm GHz}$
light curve will be closer to a pure gravitational lensing event.  We
fit the $8\,{\rm GHz}$ light curve taking as a model lens
a constant surface density disc with
$R=2.5\,\x0=6\times 10^{14}\,{\rm cm}$ and $M\simeq 3\, M_\odot\d101\v2202$
The predicted light curve is shown by the dashed lines in Figs.~2a and 2b.
As predicted, the fit is much better in the shorter wavelength channel.

F2 estimate that there are roughly $250-450$ clouds per ${\rm arcsec}^2$.
This is comparable to but somewhat higher than the number density of
gravitational microlenses that would be inferred from the MACHO and EROS
results.  Our suggestion is therefore that ESEs and the MACHO and EROS
events are caused by related objects but {\it not} that all partially
ionized clouds
capable of causing an ESE are
dense enough to cause gravitational amplification.

The question remains as to whether dark baryonic clouds are theoretically
plausible and, more to the point, whether such objects are preferred over low
mass stars and brown dwarfs in the evolution from  primordial densities.
New physics, namely different stellar mass functions for the disc and
halo, would be required to explain a large population of brown dwarfs or
low mass stars in the halo.
This may not be so difficult to imagine
as halo material is lower in metals and dust which are important for cooling.
Moreover, magnetic fields, important for removing angular momentum from a
rotating and collapsing cloud, may be absent.  The conjecture
that baryons in the halo end up in clouds rather than compact objects
takes this one step further by suggesting that star
formation, at least in its final phases, is
far less efficient in the halo than in the disc.

Consider first the simple case of an isolated cloud.  Recall
that our prototype lens has $M\simeq 0.1\,M_\odot$, $R\simeq 10^{14}\, {\rm
cm}$
and $\sigma\simeq 2\, {\rm km/sec}$.  For an isothermal sphere, the
corresponding
temperature is $T\simeq 500\,^\circ K$ and the mass of such an object is
roughly
the Jeans mass $M_J=\left (\pi kT/Gm_p\mu\right )^{3/2}\rho^{-1/2}$ where
$m_p$ is the mass of the proton and $\mu$ is the mean
molecular weight.  Objects more massive than the Jeans mass
might persist if they are supported in two dimensions by rotation provided
that the Toomre stability
condition $Q\equiv v_s\kappa/\pi G\Sigma\ga 3$ (Binney and Tremaine, 1987,
Chpt. 6)
is satisfied.  In this expression,
$v_s$ is the sound speed and $\kappa$ is the epicyclic
frequency.  Suppose we have an isothermal disc whose characteristic
thickness $z_0$ is much smaller than its radius.  It is straightforward
to show that $\sigma_z^2=2\pi z_0G\Sigma$ where $\sigma_z$ is the
velocity dispersion perpendicular to the plane of the disc
(Binney and Tremaine 1987, p.\. 282 and Spitzer 1942).
Assuming $\Sigma=\Sigma_{\rm cr},~v_s=\sigma_z$, and $\kappa=v_c/R$
where $v_c$ is the circular velocity at the edge of the disc, the Toomre
stability
criterion becomes
\be
\frac{v_c}{c}\left (\frac{z_0}{R}\right )^{1/2}\ga 3\times 10^{-5}
\ee
where we have set $D=10\,{\rm kpc}$ and $R=2.5\times 10^{13}\,{\rm cm}$.
The above condition can be easily satisfied though one should keep in mind that
there is no guarantee such
an object will be stable to very high order, nonaxially-symmetric
instabilities if the Jeans mass becomes very small with respect to the total
mass.

A Jeans mass object which radiates a significant
fraction of its energy will of course contract.
For a pure hydrogen and helium cloud, radiation is due mainly to
molecular hydrogen transitions.
Palla, Salpeter, and Stahler (1983) study the evolution of just such a cloud.
They find that
the high density phase of the cloud's evolution is roughly independent of the
starting
conditions and is greatly affected
by the formation of $H_2$ through 3-body reactions.  Their Fig. 3 shows that
the
temperature, density, and Jeans mass all pass through
typical values for the clouds we require.
Nevertheless, by their calculations, the clouds continue to evolve through
this region on the free-fall time-scale due to the overlap in density of the
molecular cooling regime with the regime of
collisional ionization and hence of bremsstrahlung cooling.  The calculation
is however dynamically naive (rotation, inhomogeneity and
magnetic fields are neglected) and it does not seem impossible for this
overlap to be reduced leaving a phase with
very low fractional levels of both $H_2$ abundance and ionization.
The largely $HI$ cloud that results cools very inefficiently and might be
quasi-stable.

Another, perhaps more
interesting possibility is that there is an ensemble of clouds having
a supersonic velocity dispersion.  If the collision time is shorter than the
cooling time, shock heating can dissociate the $H_2$ well
before the occurrence of collisional ionization (Palla, Salpeter, and Stahler
1983).

The picture of an ensemble of clouds is similar, in some respects, to the
ideas proposed by PCM and PC.  These authors suggest that dark matter around
spiral galaxies is in the form of cold $H_2$ gas in a fractal structure.
The gas is assumed to be in thermal equilibrium with the microwave background
and therefore has a temperature of $3 ^\circ K$.
The smallest indivisible elements in the fractal distribution, called
clumpuscules, are
set by equating the free-fall time with the Kelvin-Helmholtz timescale
(Rees 1976) and have a mass $M\simeq 4\times 10^{-3} M_\odot
\left (T/^\circ K\right )^{1/4}\mu^{-9/4}\simeq 10^{-3} M_\odot$
and radius $R\simeq 2.3\times 10^{15} {\rm ~cm}\left (T/ ^\circ K\right
)^{-3/4}
\mu^{-5/4}\simeq 10^{15}\,{\rm cm}$.  The largest structures in the fractal are
of the scale of molecular clouds with
$M=10^6~M_\odot$ and $R\simeq 30~{\rm pc}$.  PC therefore find a fractal
dimension $d\equiv\log{(M/M_0)}/\log{(R/R_0)}\simeq 1.7$.  They argue that
the structures along the fractal will be in statistical equilibrium
with coalescence, fragmentation, and evaporation being the main processes.
Collisions occur often enough to disrupt the clumpuscules and
prevent collapse and the formation of brown dwarfs or low mass stars.
We imagine a similar picture but with a fragmentation temperature
of $500^\circ K$ and $d\simeq 1.1$.  The resulting fractal structure would then
contain objects capable of explaining the MACHO/EROS results.  Such a fractal
structure would {\it not} have a very large sky covering factor.  PC
(Fig. 8) show for example, that half of the mass in the ensemble would
be in $0.3\%$ of the area on the sky for $d=1.5$ assuming that elements
of the distribution are truncated isothermal spheres.  The covering factor
is even less at $d=1.1$

Another type of fractal distribution which has a mixture of sub-clouds and
diffuse
gas on each scale was introduced as a model for molecular clouds by
Henriksen and Turner (1984), Henriksen (1986) and reviewed by Henriksen (1991).
In this model, a statistical mechanism for solving the well-known
angular momentum problem is proposed.  In this hierarchically virialized,
collisional ensemble, angular momentum is transported
from a fraction of the clouds on small scales to larger scales principally
by the action of tidal torques during collisions of the small-scale clouds.
There is also a fraction of the small-scale clouds that are spun up by
the collisional interactions, and consequently one measure of the
star formation efficiency would be the fraction that the `spun-down' clouds are
of the total.  Roughly speaking, star formation is shut off if the collision
time of
small-scale clouds, $t_c$, is small compared to the time $t_J$ for these clouds
to become Jeans unstable due to cooling and angular momentum loss.  For in
this case, a given cloud will be spun up and heated to virial temperatures by
collisions before collapsing.  The model leads to a dynamical population of
virialized clouds `rebounding' from the small scale $R$ where
$t_c(R)\ll t_J(R)$.  The smallest clouds in the ensemble may well be the
objects responsible for the MACHO and EROS events.

The fractal structures described above are an example of a simple, but
very specific way in which the clouds are distributed in mass and radius,
namely one in which there is a one-to-one relationship between mass
and radius with $M\propto R^d$.  More generally, there will be
a distribution of radii for each mass.
In any case, the implication is that there will be lensing events
of different duration times.  For distributions in which
$M\propto R^d$ with $d<2$ (which is usually the case), $R/\x0\propto
M^{\frac{2-d}{2d}}$
so that above some characteristic mass, $R\gg\x0$ and gravitational
amplification becomes negligible.  It is in the transition
region, where $R = {\it few}\times \x0$ and lensing occurs but is easily
distinguished from lensing by a compact object, that we have the best hopes
of confirming our hypothesis.  This makes the 1502+106 light curve all the
more intriguing.  It also suggests that the MACHO and EROS groups
may want to reexamine their selection criteria.  In particular,
events might be selected whose light curves are roughly independent of
frequency
but do not fit theoretical light curves calculated assuming
point mass lenses.  For example, clouds may cause lensing events that
are not time-symmetric.  Indeed, by selecting candidate events based on
light curves derived assuming point mass lenses,
we may be significantly underestimating the density
of dark baryons in the galaxy.

\bigskip

\centerline{\bf Acknowledgements}

\bigskip

It is a pleasure to acknowledge A. Babul,
R. Feidler, V. Hughes, J. Irwin, K. Lake, A. Meiksin, R. Nelson
and S. Tremaine for useful conversations.  Special thanks go to R. Feidler
for providing us with light curves for 1502-106 (F2).

\vfill
\eject
\centerline{\bf REFERENCES}
\frenchspacing
\baselineskip 12pt

\medskip

Alcock, C. et al. 1993, Nature, {\bf 365}, 621

\medskip

Aubourg, E. et al. 1993, Nature, {\bf 365}, 623

\medskip

Binney, J. \& Tremaine, S. 1987, Galactic Dynamics (Princeton, N. J.: Princeton
University Press)

\medskip

Cognard, I., Bourgois, G., Lestrade, J., Biraud, F., Aubry, D., Darchy, B.,
Drouhin, J. 1993, Nature {\bf 366}, 320

\medskip

Fiedler, R. L., Dennison, B., Johnston, K. J., \& Hewish, A. 1987, Nature
{\bf 326}, 675 (F1)

\medskip

Fiedler, R. L., Dennison, B., Johnston, K. J., Waltman, E., \& Simon, R. 1993,
Naval Research Laboratory preprint (F2)

\medskip

Genzel, R. 1992, in: The Galactic Interstellar Medium, D. Saas-Fee Advanced
Course 21,
Pfenniger, P. (ed.), Springer-Verlag, Berlin, p. 275

\medskip

Griest, K. 1991, ApJ {\bf 366}, 412

\medskip

Henriksen, R. N. \& Turner, B. E. 1984, ApJ {\bf 287}, 200

\medskip

Henriksen, R. N. 1986, ApJ {\bf 310}, 189

\medskip

Henriksen, R. N. 1991, ApJ {\bf 377}, 500

\medskip

Hogan, C. J. \& Rees, M. J. 1988, Phys. Lett. {\bf B205}, 228

\medskip

Kolb, E. W. \& Tkachev, I. I. 1993, Phys. Rev. Lett. {\bf 71}, 3051

\medskip

Leibert, J. \& Probst, R. G. 1987, ARA\&A {\bf 25}, 473

\medskip

Paczynski, B. 1986, ApJ {\bf 304}, 1

\medskip

Palla, F., Salpeter, E. E., \& Stahler, S. W. 1983, ApJ {\bf 271}, 632

\medskip

Pfenniger, D., Combes, F., \& Martinet, L. 1994, A\&A in press (PCM)

\medskip

Pfenniger, D. \& Combes, F. 1994, A\&A in press (PC)

\medskip

Rees, M. J. 1976, MNRAS {\bf 176}, 483

\medskip

Romani, R. W., Blandford, R. D., \& Cordes, J. M. 1987, Nature {\bf 328}, 324

\medskip

Schneider, P., Ehlers, J., \& Falco, E. E. 1992, Gravitational Lenses
(Berlin: Springer-Verlag)

\medskip

Silk, J. \& Stebbins, A. 1993, ApJ {\bf 411}, 439

\medskip

Spitzer, L. 1942, ApJ {\bf 95}, 329

\medskip

Walker, T. P., Steigman, G., Schramm, D. N., Olive, K. A., and Kang, H.-S.
1991, ApJ, {\bf 376}, 51

\vfill
\eject
\centerline{\bf FIGURE CAPTIONS}
\bigskip
\noindent {\bf Figure 1:}  Light curves for various extended lens models as
compared
with the light curve derived assuming a point mass lens.  In each figure, the
solid line refers to the extended lens and the dashed line refers to the point
mass
lens.  Figs. 1a,\, 1b, and 1c are for a singular isothermal sphere, a constant
density sphere, and a constant surface density disc respectively.  The curves
are
fit to give an amplification factor of 6.86 at the peak (${\rm time}=0$)
and can therefore be used to fit the single event reported by the MACHO
collaboration
(Alcock et al. 1993).  The horizontal scale is normalized so that $A=1$ at
${\rm time}=\pm T$.  For the MACHO event, $T\simeq 9\,{\rm days}$.

\bigskip

\noindent {\bf Figure 2:}  Radio frequency light curves for the quasar 1502+106
at $8.1$ and $2.7\,{\rm GHz}$.  ${\rm time}=0$ roughly corresponds to November
1985.
The dashed line is a theoretical light curve derived assuming a constant
surface density disc with radius $2.5$ times the Einstein radius (see text).
The model was chosen to fit the $8.1\,{\rm GHz}$ data where scattering by
electrons
is expected by be negligible and where the quasar can be modeled as a point
source.
\end{document}